\documentclass[a4paper,12pt]{article}
\newcommand{\beqn}{\begin{equation}}
\newcommand{\eeqn}{\end{equation}}
\newcommand{\barr}[1]{\begin{array}{#1}}
\newcommand{\earr}{\end{array}}
\newcommand{\beqna}{\begin{eqnarray}}
\newcommand{\eeqna}{\end{eqnarray}}
\newcommand{\lapprox}{\stackrel{<}{\scriptstyle \sim}}
\newcommand{\rapprox}{\stackrel{>}{\scriptstyle \sim}}
\newcommand{\cc}{$c\bar{c}$}
\def\question#1{}

\begin{document}
\title{
\begin{flushright}
\small{hep-ph/0507199}
\\ \small{LA-UR-05-5675}
\end{flushright}
\vspace{0.6cm}
\Large\bf Gluonic charmonium resonances at BaBar and Belle?}
\vskip 0.2 in
\author{Frank E. Close\thanks{\small \em E-mail: F.Close1@physics.ox.ac.uk} \\
{\small \em Department of Physics - Theoretical Physics,}\\
{\small \em University of
Oxford, 1 Keble Road, Oxford OX1 3NP, U.K. }\vspace{.3cm} \\
Philip R. Page\thanks{\small \em E-mail: prp@lanl.gov} \\
{\small \em Theoretical Division, MS B283, Los Alamos
National Laboratory,}\\
{\small \em  Los Alamos, NM 87545, U.S.A.}}
\date{}
\maketitle
\begin{abstract}{
We confront predictions for hybrid charmonium and other gluonic
excitations in the charm region with recently observed structures in
the mass range above 3 GeV. The $Y(4260)$, if resonant,
is found to agree with
expectations for hybrid charmonium.  The possibility that other
gluonic excitations may be influencing the data in this region is
discussed.
}
\end{abstract}
\bigskip

PACS number(s): 12.39.Mk
\hspace{.1cm} 13.25.Gv \hspace{.1cm} 14.40.Gx


\vspace{1.5cm}

\section{Introduction}

In a series of papers since 1995 we have defined the properties of
 gluonic hybrid charmonium and developed a strategy for producing and
 identifying such 
states~\cite{cp95,pthesis,bcs,cp96,dunietz,pss,pbar,cgodfrey}. 
In this
 paper we compare these predictions with recently observed structures
 in the mass region above 3 GeV.  We shall argue that the 
$Y(4260)$, if resonant, 
 has properties consistently in line with our historical
 predictions.

Our starting point is that four recent experiments have reported the
discovery of broad states consistent with charmonia: $Y(3940)$ seen in
$J/\psi\:\omega$~\cite{bellepsiomega}, $X(3940)$ seen in
$D^\ast\bar{D}$~\cite{belleddstar}, $\chi_{c2}(3930)$~\cite{bellechi2}
and $Y(4260)$~\cite{babar}. (The inclusion of charge-conjugated
reactions is implied throughout this paper.) Furthermore there are also
three prominent enhancements
$X$ in $e^+e^- \to J/\psi + X$~\cite{belleddstar}, which are
consistent with being the $\eta_c,\eta_c'$ and $\chi_0$.

 In this paper we address the question of whether any of these states
 may signal the excitation of gluonic degrees of freedom in the
 charmonium regime.

(i) The $Y(3940)$ has been supposed to be hybrid
charmonium~\cite{bellepsiomega}: we critically assess this claim.

(ii) By contrast, the $Y(4260)$~\cite{babar} has mass, width,
production and decay properties, all in accord with those that we
predicted historically for hybrid charmonium.

(iii) The prominent enhancements $X$ seen in $e^+e^- \to J/\psi +
X$~\cite{belleddstar}: We suggest that these states be studied further
as their production may be strongly affected by $C=+$ glueballs
predicted to occur in this range, and there are {\it prima facie}
inconsistencies with simply associating them with known \cc~ states.

We open with some brief remarks about points (iii) and then (i); the
 main thrust of this paper will be to discuss in detail the evidence
 related to the hybrid charmonium hypothesis, point (ii).

In pQCD the amplitude for $e^+ e^- \to J/\psi + c\bar{c}$ is the same
order as $e^+ e^- \to J/\psi + gg$ and has led to the suggestion
\cite{brodskyglue} that $C=+$ glueballs could be produced at a
significant level.  Although the coupling of such states to light
flavours may give them large widths, and make a simple glueball
description naive, it is nonetheless possible that their presence may
enhance the production of \cc~ states with the same $J^{PC}$.  We note
that in lattice QCD gluonic activity in the $0^{-+}$ channel is
predicted $\sim 3.6$ GeV~\cite{latticeglue},
which is potentially degenerate with the mass
for the $\eta_c(2S)$~\cite{pdg05}. Note there are potentially different
masses obtained for the state in electromagnetic
($\gamma \gamma$) and $B \to K \eta_c$
decays~\cite{pdg05}; thus it is possible
that different production mechanisms expose the presence of
non-trivial mixing between the \cc~ and gluonic sectors here.  Lattice QCD
also predicts activity in the gluonic waves $1^{++}$, $1^{-+}$, $2^{-+}$  and
$3^{++}$ in this mass region~\cite{latticeglue}.
We advocate that until careful spin-parity analysis is done,
one should be cautious about identifying
these enhancements naively with \cc~ states.

Indeed, there are already some potential problems with the specific
 fits to $e^+e^- \to J/\psi + X$ in Ref.~\cite{belleddstar}, which
 assigns the resonance bumps on the basis of the masses of states in
 the PDG~\cite{pdg}.  As a result they identify $\chi_0$ but no
 prominent $\chi_{1,2}$, though there may be room for these states in
 the small fluctuations around $3.5-3.6$ GeV in Fig. 1 of
 Ref.~\cite{belleddstar}.  The Born cross-sections containing more than
 two charged tracks are approximately for $X= \eta_c: 25.6$ fb;
 $\chi_0: 6.4$ fb; $\eta_c': 16.5$ fb~\cite{pak} and
 $X(3940): 10.6$ fb~\cite{belleddstar}.

There is no sign of the $X(3872)$; this state now appears to have
$C=+$ and be consistent with $1^{++}$~\cite{3872data}. This $J^{PC}$
was first suggested in Ref.~\cite{fcpage3872} and a dynamical picture of it
as a quasi-molecular $D^{\ast 0}\bar{D}^0$ state discussed
in Refs.~\cite{fcpage3872,swanson3872}. The suppression of this state among
prominent $C=+$ charmonium states~\cite{belleddstar} may thus be
consistent with its molecular versus simple \cc~ nature.

A natural first guess is to associate the $X/Y(3940)$ with radially
excited $2P$ charmonium.  In $\gamma\gamma$ production the radial
charmonium $\chi_2(3930)$ has now been reported~\cite{bellechi2}, which
gives a benchmark for comparing the other novel states.  The nearness
of $X(3940)$ and $Y(3940)$ to the $\chi_{c2}(3930)$ suggests that
these state(s) are consistent with radially excited P-wave charmonia
which are predicted in that mass region. In particular, decays to
$J/\psi\: \omega$ and $D^\ast\bar{D}$ with absence of $D\bar{D}$ are 
consistent
with these two states being the same and identified as being
$2P(^3P_1)$.  This assignment has the advantage that the
phase-space limited $J/\psi\: \omega$ mode is in S-wave, while 
another possibility, $0^{-+}$, does not share this feature. 

It is possible that the effect of nodes in the wavefunctions of
radially excited states could cause an accidental suppression of
$D\bar{D}$, for example in $2P(^3P_0)$ and confuse the identification
of excited charmonium states; this seems unlikely if the results of
Ref.~\cite{bgs} are a guide.  However, the absence of a prominent
$\chi_1$ state in the data advises caution in identifying the
prominent $X/Y(3940)$ as solely the radially excited $2P(^3P_1)$
state. If one interprets $e^+e^- \to J/\psi + X$ as a measure of the
cross section for $X\; (C=+)$, then the meson pair production thresholds
with $C=+$ are opening in relative S-waves in this mass region and so
some of the structure may reflect the opening of such channels rather
than being simply resonant.  Angular distributions of, for example,
$D\bar{D},D^\ast\bar{D}$ and $D^\ast \bar{D}^\ast$ 
should be investigated to establish if there
are specific resonances or alternatively threshold effects driving the
enhancement.

Such information already exists qualitatively and can help to
constrain interpretations.  If the $3940$ MeV enhancement consists of a
single state, then the observation of significant $D^\ast\bar{D}$ in the decay
of $X(3940)$ suggests that this state is not simply a gluonic hybrid
charmonium~\cite{cp95}.  However, the branching ratios into $D^\ast
\bar{D}:\: J/\psi\:\omega$ need to be established; if the 
$D^\ast\bar{D}$ is small,
then hybrid charmonium may be relevant.  The mass also is low compared
to that predicted for hybrid charmonia which are more generally
expected to be at $\sim 4.2$ GeV~\cite{bcs,ukqcd} unless, as we discuss
later, there are significant $J^{PC}$ dependent mass
shifts.

While the interpretation of these states may depend rather critically
on first establishing their $J^{PC}$, the appearance of a further
state, $Y(4260)$ with $J^{PC}=1^{--}$ has properties that appear
uniformly to be consistent with those predicted earlier for hybrid
charmonium. We now assess the experimental information about
this state.

\section{Experimental information}

The BaBar collaboration recently observed a new structure at
$4259 \pm 8^{+2}_{-6}$ MeV with a width of $88 \pm 23^{+6}_{-4}$
MeV and a significance greater than $8\sigma$~\cite{babar}.
The structure is known to be produced in initial state radiation
from $e^+e^-$ collisions and hence to have
$J^{PC}=1^{--}$. It is seen decaying to $J/\psi\:\pi^+\pi^-$ and

\beqn\label{babrel}
\Gamma(Y(4260)\rightarrow e^+e^-)\; {\cal B}(Y(4260)\rightarrow
J/\psi\:\pi^+\pi^-) = 5.5\pm1.0^{+0.8}_{-0.7}\; eV.
\eeqn

There are several consequences of the experimental work that are worth
noting, and which align themselves most naturally with a hybrid
charmonium interpretation.

{\it The mass coincides with the $D_1(2420) \bar{D}$ threshold:}
The state can couple to $D_1(2420) \bar{D}$, and related thresholds to be 
discussed later, in S-wave. The $D_1(2420) \bar{D}$ thresholds
are at $4287$ MeV ($D_1(2420)^o \bar{D}^o$) and $4296$ MeV
($D_1(2420)^\pm D^\mp$)~\cite{pdg05}. At an S-wave threshold, re-scattering
effects may drive the $J/\psi\:\pi^+\pi^-$ signal.
A resonance above $4.26$ GeV, which couples strongly to
$D_1\bar{D}$, can through re-scattering give an enhancement in 
$J/\psi\: \pi \pi$ at the $D_1\bar{D}$
threshold (for example see Ref.~\cite{donnpage}), 
in which case the true mass of $Y(4260)$ could be
$O(100)$ MeV above $4.26$ GeV. 
It is even possible for such a phenomenon to occur without
any resonance. Therefore
it is important to establish that $Y(4260)$ is resonant and not
a threshold effect. With these caveats, we shall now analyze
for the case where $Y(4260)$ is resonant.

{\it The decay modes $J/\psi\: \sigma$, $J/\psi\: f/a_0(980)$ appear to
dominate:} An S-wave phase space model of the three-body decay
$J/\psi\:\pi^+\pi^-$~\cite{babar} does not appear consistent with
the data (Fig. 3, Ref.~\cite{babar}). On the other hand, two-body
decay, which usually dominates three-body decay,
would easily explain the data, which are consistent with $\pi^+\pi^-$
peaks at the $\sigma$ and $f_0(980)/a_0(980)$ masses. These mesons
are the only ones in the mass region $0.3-1.0$ GeV displayed~\cite{babar}
with $C=+$, as required by C-parity conservation.
The mode $J/\psi\: K\bar{K}$ should be searched for as the strong coupling of
$f/a_0(980)$ to $K\bar{K}$ should manifest itself at the $K\bar{K}$ threshold
if these states are important in the decay.

{\it $\Gamma(Y(4260)\rightarrow e^+e^-)$ is much smaller than
all other $1^{--}$ charmonia:} Noting that the cross-section
$\sigma(e^+e^- \rightarrow Y \rightarrow X)$ into final state
$X$ is proportional to
$\Gamma(Y \rightarrow e^+e^-) {\cal B}(Y \rightarrow X)$,

\beqn
\frac{\Gamma(Y \rightarrow e^+e^-) {\cal B}(Y \rightarrow
\mbox{hadrons})}{\Gamma(Y \rightarrow e^+e^-) {\cal B}(Y \rightarrow
J/\psi\:\pi^+\pi^-)} =
\frac{\sigma(e^+e^- \rightarrow Y \rightarrow
\mbox{hadrons})}{\sigma(e^+e^- \rightarrow Y \rightarrow J/\psi\:\pi^+\pi^-)},
\eeqn
and using Eq.~\ref{babrel}, $\sigma(e^+e^- \rightarrow Y \rightarrow$
hadrons$) \lapprox 4\% \times 14.2$ nb~\cite{babar,pdg}, and
$\sigma(e^+e^- \rightarrow Y \rightarrow J/\psi\:\pi^+\pi^-) \approx 50$
pb~\cite{babar}, it follows that

\beqn\label{eewidth} 5.5\pm 1.3\; eV \leq \Gamma(Y(4260) \rightarrow
e^+e^-) \lapprox 62\pm 15 \; eV, \eeqn using that ${\cal B}(Y
\rightarrow$ hadrons$)$ is very near to unity. (The lower bound on the
width is obtained from Eq.~\ref{babrel}.)  This $e^+e^-$ width is at
least a factor of $4$ smaller than that of the established $1^{--}$
charmonium with the smallest width, the $\psi(3770)$~\cite{pdg}.
However, unmixed radially excited D-wave (2D) $c\bar{c}$ states can
have widths consistent with Eq.~\ref{eewidth}, as their widths in
potential models are typically 64 times lower than 3S
states~\cite{tornqvist}.  The experimental width only just overlaps
with that in a four-quark interpretation of $Y(4260)$, which predicted
that it should be $50-500$ eV~\cite{maiani}.

{\it $\Gamma(Y(4260)\rightarrow J/\psi\:\pi^+\pi^-)$ is much larger
than all $1^{--}$ charmonia:}
Using Eqs.~\ref{babrel} and~\ref{eewidth} it is immediate that

\beqn\label{pipi}
{\cal B}(Y(4260)\rightarrow J/\psi\:\pi^+\pi^-) \rapprox 8.8 \% ;
\hspace{0.2cm} \Gamma(Y(4260)\rightarrow J/\psi\:\pi^+\pi^-) \rapprox
7.7\pm 2.1 \; MeV.
\eeqn
This is much larger than $\Gamma(\psi(3770)\rightarrow
J/\psi\:\pi^+\pi^-)$ which is in the $80-90$ keV range~\cite{pdg,bes}.
It is also much larger than $\Gamma(\psi(4040); \psi(4160); \psi(4415)
\rightarrow J/\psi\:\pi^+\pi^-)$, as is now shown. The
$Y(4260)$ is seen in the BaBar experiment and
$\psi(4040), \psi(4160)$ and $\psi(4415)$ not~\cite{babar}.
Using a ball-park estimate that the error bars can mask the latter resonances
if their cross-section is four times smaller than $Y(4260)$
(Fig. 1, Ref.~\cite{babar}), and noting that
$\sigma(e^+e^- \rightarrow X \rightarrow J/\psi\:\pi^+\pi^-)$ into 
intermediate
state $X$ is proportional to
$\Gamma(X \rightarrow e^+e^-) {\cal B}(X \rightarrow J/\psi\:\pi^+\pi^-)$, 
we have

\beqn
{\cal B}(\psi'\rightarrow J/\psi\:\pi^+\pi^-) \lapprox
\frac{\Gamma(Y \rightarrow e^+e^-) {\cal B}(Y \rightarrow
J/\psi\:\pi^+\pi^-)}{4\, \Gamma(\psi' \rightarrow e^+e^-)} ,
\eeqn
where $\psi'$ denotes any of $\psi(4040), \psi(4160)$ or $\psi(4415)$.
Together with Eq.~\ref{babrel} this can be used to calculate a bound on
the branching ratio of each $\psi'$. Translating into widths~\cite{pdg}

\beqn
\Gamma(\psi(4040);\psi(4160);\psi(4415)\rightarrow J/\psi\:\pi^+\pi^-)
\lapprox 100 \pm 30, 140 \pm 60, 130 \pm 60 \; keV.
\eeqn

The simplest interpretation of this is that the $J/\psi\: \pi\pi$ is not
due to disconnected $J/\psi\: gg$ diagrams but instead involves some
strong affinity. This could be due to a four-quark
interpretation~\cite{maiani} or due to intrinsic gluonic excitation in
the initial state, as will be discussed below.

\section{$Y(4260)$ as hybrid charmonium}

Lattice QCD inspired the flux-tube model of mesons~\cite{ipaton}, which
has been used to predict observables that, at the time, were beyond
the bounds of lattice computation but which have subsequently been
largely confirmed and extended by these more fundamental techniques.
In particular and of relevance to the present discussion, we cite the
early prediction of exotic $J^{PC}$ states for both light and heavy
flavours, whose masses and spin dependent mass splittings are now
being confirmed by lattice computations. The detailed production and
decay signatures for hybrid states are still largely beyond the bounds
of lattice QCD, and for these we are still restricted to the model.
In due course we anticipate that these results will be tested by the
lattice. In the meantime they are arguably the nearest we have and it
is on the basis of their implications that we proceed to examine the
$Y(4260)$.

 The eight low-lying hybrid charmonium states ($c\bar{c}g$) were
predicted in the flux-tube model to occur at $4.1-4.2$ GeV~\cite{bcs},
and in UKQCD's quenched lattice QCD calculation with infinitely heavy
quarks to be $4.04\pm 0.03$ GeV (with un-quenching estimated to raise
the mass by $0.15$ GeV)~\cite{ukqcd}.  The splittings of $c\bar{c}g$
from the above spin-average was predicted model-dependently for long
distance (Thomas precession) interactions in the flux-tube
model~\cite{merlin}, and for short distance
(vector-one-gluon-exchange) interactions in cavity
QCD~\cite{pthesis,pbar}. For the $1^{--}$ state the long and short
distance splittings are respectively $0$ MeV and $60$ MeV.

These mass predictions are very much in accord with the $Y(4260)$ and
somewhat removed from those for $X/Y(3940)$.

A lattice inspired flux-tube model showed that the decays of hybrid
mesons, at least with exotic $J^{PC}$, are suppressed to pairs of
ground state $1S$ conventional mesons~\cite{ipaton,ikp}.  This was extended
to all $J^{PC}$, for light or heavy flavours in Ref.~\cite{cp95}.
A similar selection rule was found in constituent gluon 
models~\cite{pene} and later in QCD sum rules~\cite{zhusel}, and
their common quark model origin is now understood~\cite{pagesel}.
It was further shown that these selection rules for light flavoured
hybrids are only approximate, but that they become very strong for
$c\bar{c}$~\cite{cp95,pthesis}.  This implied that decays into
$D\bar{D},\: D_s\bar{D}_s,\: D^\ast \bar{D}^\ast$ and $D_s^\ast
\bar{D}_s^\ast$ are suppressed whereas $D^\ast\bar{D}$ and
$D_s^\ast\bar{D}_s$ are small, and $D^{\ast\ast}\bar{D}$, if above
threshold, would dominate. (P-wave charmonia are
denoted by $D^{\ast\ast}$).  As $c\bar{c}g$ is predicted around the
vicinity of $D^{\ast\ast}\bar{D}$ threshold, the opportunity for
anomalous branching ratios in these different classes was proposed as
a sharp signature~\cite{cp95,bcs}. (To the best of our knowledge
Ref.~\cite{cp95} was the first paper to propose such a distinctive
signature for hybrid charmonium.)

It has become increasingly clear recently that there is an affinity
for states that couple in S-wave to hadrons, to be attracted to the
threshold for such channels~\cite{fcichep}. The hybrid candidate
$1^{--}$ appearing at the S-wave $D_1(2420)\bar{D}$ is thus interesting.

 More recently the signatures for hybrid charmonia were expanded to
note the critical region around $D^{\ast\ast}\bar{D}$ threshold as a
divide between narrow states with sizable branching ratio into
$c\bar{c}\; +$ light hadrons and those above where the anomalous
branching ratios would be the characteristic
feature~\cite{dunietz,pbar,cgodfrey}.  Here 
widths of order 10 MeV were anticipated around the threshold.  It
was suggested to look in $e^+e^-$ annihilation in the region
immediately above charm threshold for state(s) showing such anomalous
branching ratios~\cite{cgodfrey}. The leptonic couplings
 to $e^+e^-,\; \mu^+\mu^-$ and $\tau^+\tau^-$  were expected
to be suppressed~\cite{ono} 
(smaller than radial S-wave $c\bar{c}$ but larger
than D-wave $c\bar{c}$, but with some inhibition due to the fact that
in hybrid vector mesons spins are coupled to the $S=0$, whose coupling
to the photon is disfavoured~\cite{cgodfrey}).  Even stronger
suppression obtains for $\gamma\gamma$ couplings~\cite{photon}.

  Small conventional charmonium mixing with $c\bar{c}g$ or a glueball
is expected. The latter is due to the penalty incurrent by the creation
of a $c\bar{c}$ pair, and the former is due to the heaviness of the
charm quark which enable a Born-Oppenheimer approximation, separating
conventional and hybrid charmonia by virtue of their orthogonal gluonic
wave functions~\cite{pbar}.  
However, for $1^{--}$ hybrids, there is the possibility
of substantial mixing with the radially and orbitally excited $c\bar{c}$
if mass degenerate: It was noted that hybrid charmonia with
$1^{--}$ can in principle mix~\cite{cp96,gerasimov} 
with radially excited $c\bar{c}$ states
and a specific example was discussed of what would occur if the hybrid
mass is $\sim 4.1$ GeV~\cite{cp96}.

 The discovery of $Y(4260)$ signals degrees of freedom beyond
conventional $c\bar{c}$. This is because the only such $1^{--}$
expected up to $4.4$ GeV are $1S,\; 2S,\; 1D,\; 3S,\; 2D$ and 
$4S$~\cite{tornqvist}, and there
are already established candidates for these states. Thus even in the
case of mixing, the existence of $Y(4260)$ hints that more than
conventional $c\bar{c}$ is needed.  

We now consider tests and implications of the idea that $Y(4260)$
signals the onset of hybrid charmonium.  We describe these below,
compare with the unmixed hybrid charmonium hypothesis and propose
further tests.

\section{Implications of hybrid charmonium}

There are several of the theoretical expectations already given for
$c\bar{c}g$ that are born out by $Y(4260)$: (1) Its mass is
tantalizingly close to the prediction for the lightest hybrid
charmonia; (2) The expectation that the $e^+e^-$ width should be
smaller than for S-wave $c\bar{c}$ is consistent with
Eq.~\ref{eewidth};
(3) The predicted affinity of hybrids to $D^{\ast\ast}\bar{D}$
could be related to the appearance of the state near the
$D^{\ast\ast}\bar{D}$ threshold. The formation of $D^{\ast\ast}\bar{D}$
at rest may lead to significant re-scattering into $J/\psi\:\pi^+\pi^-$,
which would feed the large signal (Eq.~\ref{pipi}).

Quenched lattice QCD indicates that the $c\bar{c}g$
$1^{--},\; (0,1,2)^{-+}$ are less massive than
$1^{++},\; (0,1,2)^{+-}$~\cite{juge}. The spin splitting for this lower
set of hybrids in quenched lattice NRQCD is
$0^{-+}<1^{-+}<1^{--}<2^{-+}$~\cite{drum}, at least for $b\bar{b}g$.
This agrees with the ordering found in the model-dependent
calculations for $q\bar{q}g$~\cite{bcd} in the specific case of
$c\bar{c}g$~\cite{pthesis,pbar,merlin}. For $b\bar{b}g$ lattice QCD
predict substantial splittings $\sim 100$ MeV or greater~\cite{drum},
which become even larger in the model-dependent calculations for
$c\bar{c}g$~\cite{pthesis,pbar,merlin}.  Theory hence strongly
indicates that if $Y(4260)$ is $c\bar{c}g$, and the splittings are not
due to mixing or coupled channel effects, then the $J^{PC}$ exotic
$1^{-+}$ and non-exotic $0^{-+}$ $c\bar{c}g$ are below
$D^{\ast\ast}\bar{D}$ threshold, making them narrow by virtue of the
selection rules. The $1^{-+}$ decay modes~\cite{dunietz} and branching
ratios~\cite{cgodfrey} have extensively been discussed.

 The nearness of $Y(4260)$ to the $D_1(2420) \bar{D}$ threshold, and
to the $D_1' \bar{D}$ threshold, with the broad $D_1'$ found at a mass
of $\sim 2427$ MeV and width $\sim 384$ MeV~\cite{dmass}, indicate
that these states are formed at rest. Also, these are the lowest open
charm thresholds that can couple to $1^{--}$ in S-wave (together with
$D_0 \bar{D}^\ast$, where the $D_0$ mass $\sim 2308$ MeV and 
width $\sim 276$ MeV~\cite{dmass}).  Flux-tube model predictions are
that the D-wave couplings of $1^{--}\; c\bar{c} g$ to the $1^{+}$ and
$2^{+}$ $D^{\ast\ast}$ are small~\cite{cp95,pthesis,pss}; and there is
disagreement between various versions of the model on whether the
S-wave couplings to the two $1^{+}$ states are large. If these
couplings are in fact substantial, the nearness of $Y(4260)$ to the
thresholds may not be coincidental, because coupled channel effects
could shift the mass of the states nearer to a threshold that it
strongly couples to; and it would experience a corresponding
enhancement in its wave function. The broadness of $Y(4260)$ also
implies that its decay to $D_1(2420) \bar{D},\; D_1' \bar{D}$ and
$D_0(2308)\bar{D}^\ast$ which feed down to $D^\ast \bar{D} \pi$ and $D
\bar{D} \pi$~\cite{cswanson} would be allowed by phase space and
should be searched for to ascertain a significant coupling to
$D^{\ast\ast}$.

 Flux-tube model width predictions for other charm modes
are $1-8$ MeV for $D^\ast \bar{D}$~\cite{pss},
with $D\bar{D},\: D_s\bar{D}_s,\:
D^\ast \bar{D}^\ast$ and $D_s^\ast \bar{D}_s^\ast$ even more suppressed.
 Thus a small $D\bar{D}$ and $D_s\bar{D}_s$ mode could single out the hybrid
interpretation. 
The hybrid decay pattern is very different from the $c\bar{s}s\bar{c}$
four-quark interpretation for $Y(4260)$ which decays predominantly in
$D_s \bar{D}_s$~\cite{maiani}. Thus a search for the latter channel,
or limit on its coupling, could be a
significant discriminator for the nature of the $Y(4260)$.

\section{Experimental searches and production}

 It is possible that $Y(4260)$ is not a resonance, but reflects the
opening of the $D_1(2420)\bar{D},\; D_1'\bar{D}$ and $D_0(2308)\bar{D}^\ast$
thresholds. The reason is that this is the lowest energy at which
open charm thresholds can couple to $e^+e^-$ ($1^{--}$) in 
S-wave. Thus there is the possibility that BaBar~\cite{babar} 
is observing the process 
$e^+e^- \to D^{\ast\ast} \bar{D} \to J/\psi\: \pi^+\pi^-$, where 
$J/\psi\: \pi^+\pi^-$ is produced by re-scattering.
This could occur without a resonance, or with a resonance,
as follows.
The essential ingredients are (i) the presence of a 
non-resonant background (in this case $J/\psi\: \pi\pi$);
(ii) a resonance which strongly couples to a channel 
(in this case $D^{\ast\ast}\bar{D}$); (iii) rescattering between 
the latter channel and the background. An example involving light 
quarks and an
earlier claimed signal for a hybrid meson 
(in that case the $1^{-+}$) was discussed in Ref.~\cite{donnpage};
a specifc model of rescattering involving charmonium was applied 
to the $X(3872)$ in Ref.~\cite{swanson3872}.
Hence the resonant
nature of $Y(4260)$ should be confirmed. If a similar re-scattering
effect occurs
at the $D_{s1}\bar{D}_s,\; D_{s1}'\bar{D}_s,\; D_{s0}'\bar{D}_s^\ast$ and
$D_{s2}'\bar{D}_s^\ast$ thresholds then this could be investigated
in $J/\psi\: K\bar{K}$. If Fig. 1 of Ref.~\cite{babar} shows further 
structure beyond the $Y(4260)$ enhancement 
this may be due to the $D_s^{\ast\ast} \bar{D}_s$ re-scattering in 
$J/\psi\: f_0(980)$ which yields a $J/\psi\: \pi^+\pi^-$ signal.

The nearness and S-wave coupling of $Y(4260)$ to specifically
the $D_1(2420)^0 D^0$ threshold, and also the
$D_1(2420)^\pm D^\mp$ threshold, together
with the sizable width of the state, lead to the expectation that mixing
with both thresholds will be similar, and that the charmonium nature
of $Y(4260)$ should imply that it is dominantly $I=0$, as will be
assumed in the remainder of this discussion. This can be established
by searching for the isospin violating decays $J/\psi\:\pi^0$ and 
$\pi^+\pi^-$.

If either the model dependent spin splittings are a guide, or if the
states are attracted towards S-wave thresholds, then we would expect
that the $Y(4260)$ as vector hybrid $\psi_g$ states will imply that
the $0^{-+}$ $\eta_{cg}$ and exotic $1^{-+}$ will be at or below $4.3$
GeV. The analyses of Refs.~\cite{pss,cgodfrey} then imply the following:

(i) Any decays into the disfavoured $D^\ast\bar{D}$ channel will be in the
ratios $1^{-+}: \psi_g: \eta_{cg} = 1:2:4$ apart from phase space
effects.

(ii) $\Gamma(1^{-+} \to D_1\bar{D}) > \Gamma(\psi_g \to D_1\bar{D}$).

(iii) $\eta_{cg} \to D_0\bar{D}$ may be significant due to the broad width
of the $D_0$. Even if this is kinematically suppressed, significant
re-scattering may result into $J/\psi\: \omega$~\cite{swanson3872,cst}.
Hence the possibility that $X/Y(3940)$ contains $\eta_{cg}$ may be
realized; establishing the $J^{PC}$ of the 3940 MeV structure(s) is thus
important.

(iv) $\eta_{cg} \to \eta_c \pi\pi$ may be anticipated~\cite{cgodfrey}
and $\eta_{cg} \to \eta_c f_0(980)$ may be a significant
contributor. To this end, a search for $\eta_{cg}(3940) \to \eta_c
K\bar{K}$ is also merited.  The $J/\psi\; \{\omega,\phi\}$ mode may be 
experimentally most tractable. 

 It is singular that apart from the $\psi(2S)$, no other states
are visible in the BaBar data~\cite{babar} until the $Y(4260)$.  
Given that its $e^+e^-$
coupling is small, this observation suggests that it is the
special affinity of this state for the $J/\psi\: \pi\pi$ channel that
gives its visibility (Eqs.~\ref{eewidth}-\ref{pipi}). 
The possible decays of $Y(4260)$ are listed in Table 1.  
A further
test for the $\psi_g$ interpretation of $Y(4260)$ would be that
 $\psi_g \to \{\sigma,\eta\} \; h_c$ 
could be significant. This would arise
if the decay was driven by flux-tube de-excitation, with quark spin
conservation.  Such a mechanism is expected in the model~\cite{cp95},
though its strength is currently unquantifiable; there are suggestions
from lattice QCD that such de-excitation modes may be significant for
heavy flavours~\cite{mcn}. This particular mode could be detected by
the isospin violating mode of the $h_c \to J/\psi\: \pi$.

\begin{table}
\begin{center}
\caption{Possible two-body hadronic decay modes of $Y(4260)$,
assuming that it has $I=0$. 
Open charm modes may be suppressed by a selection
rule discussed in the text. Hidden charm modes to
low-lying charmonia are listed. For these modes,
the charmonia tend to have the same $C$ as that of the parent
$c\bar{c}g$, since, barring
non-perturbative effects, two gluons $C=+$ are
emitted in the lowest order process~\protect\cite{dunietz}. 
Electromagnetic modes
like $\{\eta_c,\eta_c(2S),\chi_{c\{0,1,2\}},h_c,X(3872)\}\gamma$
are expected to be small. Light hadron modes are restricted
to hadrons up to the $\phi$ mass.
}
\label{t1}
\vspace{0.2in}
\begin{tabular}{|c|c|c|}
\hline
Open charm & Hidden charm & Light hadrons \\
\hline
\hline
$D\bar{D};\: D_s\bar{D}_s$ &
$\eta_c\{\omega,\phi,h_1\}$ &
$\eta^{(')}\{\omega,\phi\};\; \rho\pi;\; a_0(980)\rho$ \\
$D^\ast \bar{D},\: D_s^\ast \bar{D}_s$ &
$J/\psi\{\sigma,f_0(980),\eta^{(')}\}$&
$\{\sigma,f_0(980)\}\{\omega,\phi\}$\\
$D^\ast \bar{D}^\ast;\: D_s^\ast \bar{D}_s^\ast$ &
$\psi(2S)\{\sigma,\eta\}$&
$\{K,K^\ast\}\bar{K};\; \{\kappa,K^\ast\}\bar{\kappa}$\\
$D_1(2420) \bar{D};\: D_1' \bar{D}$ &
$\chi_{c0}\omega;\: h_c\{\sigma,\eta\} $&
$K^\ast\bar{K}^\ast;\; p\bar{p},p\bar{n},n\bar{n} $\\
\hline
\end{tabular}
\end{center}
\end{table}

  A search for $\psi_g\to J/\psi\:\pi^+\pi^-$ 
at Belle and BaBar in $B\to K\psi_g$ should
be fruitful~\cite{dunietz}, 
even though the small $e^+e^-$ coupling of $\psi_g$
suggests that its wave function at the origin is tiny. 
Production in $p\bar{p}$ annihilation in the formation process
$p\bar{p}\to \psi_g$ is also feasible at future colliders.

If the $Y(4260)$ is $\psi_g$, then the radiative transition $\psi_g
\to 1^{-+} \gamma$, though tiny, may reveal the exotic hybrid
charmonium~\cite{cp96}. 
 The decay $1^{-+} \to \chi_{\{0,1,2\}} \sigma$ with
$\sigma \to (\pi \pi)_S$ should be an excellent search 
mode~\cite{dunietz}
and is predicted to be large~\cite{mcn}, although 
the $J/\psi\; \{\omega,\phi\}$ mode may be 
most tractable experimentally. 
However, given the small $e^+e^-$
width of the $Y(4260)$, this may require a dedicated search at BES or
CLEO$_c$. An exciting possibility is that $e^+e^- \to J/\psi + X$ may
reveal the $1^{-+}$ in the $X$ around or above 4 GeV~\cite{dude}.

\vskip 0.2in

While this work was in preparation, a discussion of the interpretations
for $Y(4260)$ suggested that the hybrid charmonium assignment
is favored~\cite{pene,zhu}. 
We thank S.J. Brodsky, J.J. Dudek and S.-L. Zhu for discussions.

\vskip 0.2in

{\it Note added:} After this work was completed, 
an enhancement consistent with $Y(4260)$ was observed in 
$B^- \to J/\psi\: \pi^+ \pi^- K^-$~\cite{babar1}. Such a search was
suggested earlier in this paper.


\begin{thebibliography}{9}

\bibitem{cp95} F.E. Close, and P.R. Page, Nucl. Phys. B {\bf 443}, 233
(1995).

\bibitem{pthesis} P.R. Page, D.Phil. thesis, Univ. of Oxford,
unpublished (1995).

\bibitem{bcs} T. Barnes, F.E. Close, and E.S. Swanson, Phys. Rev.
D {\bf 52}, 5242 (1995).

\bibitem{cp96} F.E. Close, and P.R. Page, Phys. Lett. B {\bf 366}, 323 (1996).

\bibitem{dunietz} F.E. Close, I. Dunietz, P.R. Page, S. Veseli, and
H. Yamamoto, Phys. Rev. D {\bf 57}, 5653 (1998).

\bibitem{pss} P.R. Page, E.S. Swanson, and A.P. Szczepaniak,
Phys. Rev. D {\bf 59}, 034016 (1999).

\bibitem{pbar} P.R. Page, Proc. of ``Workshop on a Low-Energy
$\bar{p}$ Storage Ring'', 3-5 August 2000, Chicago, IL, p. 55, hep-ph/0107016.

\bibitem{cgodfrey} F.E. Close, and S. Godfrey, Phys. Lett. B
{\bf 574}, 210 (2003).

\bibitem{bellepsiomega} Belle Collaboration, S.K. Choi {\it et al.},
Phys. Rev. Lett. {\bf 94}, 182002 (2005).

\bibitem{belleddstar} Belle Collaboration, K. Abe {\it et al.},
hep-ex/0507019.

\bibitem{bellechi2} Belle Collaboration, K. Abe {\it et al.},
hep-ex/0507033.

\bibitem{babar} BaBar Collaboration, B. Aubert {\it et al.}, hep-ex/0506081.

\bibitem{brodskyglue} S.J. Brodsky, A. Goldhaber, and  J. Lee,
Phys. Rev. Lett. {\bf 91}, 112001 (2003);
S. Dulat {\it et al.}, Phys. Lett B {\bf 594}, 118 (2004);
F.E. Close, and Q. Zhao, Phys. Rev. D {\bf 71}, 094022 (2005).

\bibitem{latticeglue} UKQCD Collaboration, G. Bali {\it et al.},
Phys. Lett. B {\bf 309}, 378 (1993);
C. Morningstar, and M.J. Peardon, Phys. Rev. D {\bf 60}, 034509 (1999).

\bibitem{pdg05} Particle Data Group
http://pdg.lbl.gov/2005/listings/mxxxcomb.html.

\bibitem{pdg} Particle Data Group, S. Eidelman {\it et al.},
Phys. Lett. B {\bf 592}, 1 (2004).

\bibitem{pak} P. Pakhlov, hep-ex/0412041; Belle Collaboration,
K. Abe {\it et al.}, Phys. Rev. D {\bf 70}, 071102 (2004).

\bibitem{3872data} CDF Collaboration, ``Measurement of the
dipion mass spectrum in $X(3872) \to J/\psi\: \pi^+\pi^-)$ decays";
http://www-cdf.fnal.gov/physics/new/ bottom/050324.blessed.X/.

\bibitem{fcpage3872} F.E. Close, and P.R. Page,
Phys. Lett. B {\bf 578}, 119 (2003); see also N.A. Tornqvist,
hep-ph/0308277 (unpublished).

\bibitem{swanson3872} E. Swanson, Phys. Lett. B {\bf 588}, 189 (2004)

\bibitem{bgs} T. Barnes, S. Godfrey, and E. Swanson, hep-ph/0505002.

\bibitem{ukqcd} P. Lacock, C. Michael, P. Boyle, and P. Rowland,
Phys. Lett. B {\bf 401}, 308 (1997).

\bibitem{donnpage} A. Donnachie, and P.R. Page, Phys. Rev. D {\bf 58},
114012 (1998).

\bibitem{tornqvist} K. Heikkil\"{a}, N.A. Tornqvist, and S. Ono,
Phys. Rev. D {\bf 29}, 110 (1984).

\bibitem{maiani} L. Maiani, F. Piccinini, A.D. Polosa, and V. Riquer,
hep-ph/0507062.

\bibitem{bes} BES Collaboration, J.Z. Bai {\it et al.}, Phys. Lett. B
{\bf 605}, 63 (2005).

\bibitem{ipaton} N. Isgur and J. Paton, Phys. Rev. D {\bf 31}, 2910 (1985).

\bibitem{merlin} J. Merlin, and J. Paton, Phys. Rev. D {\bf 35}, 1668 (1987).

\bibitem{ikp} N. Isgur, R. Kokoski and J. Paton,
Phys. Rev. Lett. {\bf 54}, 869 (1985).

\bibitem{pene} E. Kou, and O. Pene, hep-ph/0507119.

\bibitem{zhusel} S.-L. Zhu, Phys. Rev. D {\bf 60},  014008 (1999), Table VII.

\bibitem{pagesel} P.R. Page, Phys. Lett. B {\bf 402}, 183 (1997). 

\bibitem{fcichep} F.E. Close, Proc. of Hadron03, hep-ph/0311087;
Proc. of ICHEP04, hep-ph/0411396.

\bibitem{ono} S. Ono {\it et
al.}, Z. Phys. C {\bf 26}, 307 (1984); Phys. Rev. D {\bf 34}, 186 (1986).

\bibitem{photon}  P.R. Page, Nucl. Phys. B {\bf 495}, 268 (1997).

\bibitem{gerasimov} S.B. Gerasimov, Proc. of $11^{th}$
Int. Conf. on Problems of Quantum Field Theory, Dubna, Russia, 
13-17 July 1998, hep-ph/9812509. 

\bibitem{juge} J.J. Juge, J. Kuti, and C.J. Morningstar, Nucl. Phys. Proc.
Suppl. {\bf 83}, 304 (2000); T. Manke {\it et al.}, Phys. Rev.
D {\bf 57}, 3829 (1998).

\bibitem{drum} I.T. Drummond {\it et al.}, Phys. Lett. B {\bf 478}, 151
(2000); T. Manke {\it et al.}, Nucl. Phys. Proc. Suppl. {\bf 86}, 397 (2000).

\bibitem{bcd} T. Barnes and F.E. Close, Phys. Lett. B {\bf 116}, 365 (1982);
M. Chanowitz and S. Sharpe, Nucl. Phys. B {\bf 222}, 211 (1983);
T. Barnes, F.E. Close and F. de Viron, Nucl. Phys. B {\bf 224}, 241 (1983).

\bibitem{dmass} BELLE Collaboration, K. Abe {\it et al.},
Phys. Rev. D {\bf 69}, 112002 (2004).

\bibitem{cswanson} F.E. Close, and E.S. Swanson, hep-ph/0505206.

\bibitem{cst} F.E. Close, E. Swanson, and C. Thomas (in preparation)

\bibitem{mcn} UKQCD Collaboration, C. McNeile, C. Michael, and 
P. Pennanen, Phys. Rev. D {\bf 65}, 094505 (2002).

\bibitem{dude} We thank J.J. Dudek for this observation.

\bibitem{zhu} S.-L. Zhu, hep-ph/0507025.

\bibitem{babar1} BaBar Collaboration, B. Aubert {\it et al.},
hep-ex/0507090. 

\end{thebibliography}
\end{document}